\documentclass[conference]{IEEEtran}
\IEEEoverridecommandlockouts
\usepackage{cite}
\usepackage{amsmath,amssymb,amsfonts}
\usepackage{algorithmic}
\usepackage{graphicx}
\usepackage{textcomp}
\usepackage{multirow}
\usepackage{xcolor}
\usepackage{geometry}
\usepackage{arydshln}
\usepackage{paralist}
\newcommand\tstrut{\rule{0pt}{2.4ex}}

\makeatletter
\newcommand{\linebreakand}{%
  \end{@IEEEauthorhalign}
  \hfill\mbox{}\par
  \mbox{}\hfill\begin{@IEEEauthorhalign}
}
\makeatother

\def\BibTeX{{\rm B\kern-.05em{\sc i\kern-.025em b}\kern-.08em
    T\kern-.1667em\lower.7ex\hbox{E}\kern-.125emX}}
\begin{document}

\title{Radio technologies for environment-aware wireless communications 
\thanks{This research was funded by Slovenian Research Agency under grant no. P2-0016 and grant no. J2-2507.}
}

\author{\IEEEauthorblockN{1\textsuperscript{st} Tomaž Javornik}
\IEEEauthorblockA{\textit{Jožef Stefan Institute} \\ Ljubljana, Slovenia \\ tomaz.javornik@ijs.si}
\and
\IEEEauthorblockN{2\textsuperscript{nd} Andrej Hrovat} 
\IEEEauthorblockA{\textit{Jožef Stefan Institute} \\ Ljubljana, Slovenia \\ andrej.hrovat@ijs.si}
\and
\IEEEauthorblockN{3\textsuperscript{rd} Aleš Švigelj} 
\IEEEauthorblockA{\textit{Jožef Stefan Institute} \\ Ljubljana, Slovenia \\ ales.svigelj@ijs.si}
}

\maketitle

\begin{abstract}
In this paper, we critically review the potential of today's terrestrial wireless communication systems including wireless cellular technologies (GSM, UMTS, LTE, NR), wireless local area networks (WLANs), and wireless sensor networks (WSNs), for estimating channel state information (CSI), the ratio between training and information symbols and the rate of channel variation, and the potential use of CSI in environment aware wireless communications. The research reveals, that early communication systems provide means for narrowband channel estimation and the CSI is only available as channel attenuation based on signal level measurements. By increasing the spectral bandwidth of communications, the CSI is estimated in some form of channel impulse response (CIR) in almost all currently used radio technologies, but this information is generally not available outside the communication systems. Also, the CSI is estimated only for the channel with active communications. The new radio technology (NR) offers the possibility of estimating the CIR for non-active channels as well, and thus the possibility of initiating environmentally aware wireless communications.

\end{abstract}


\section{Introduction}
\label{sec:introduction}

Contemporary wireless communication systems exploit channel state information (CSI) to mitigate wireless channel impairments, increase wireless link reliability, or increase throughput. The standard approach to CSI estimation is to insert the training symbols into the transmitted signal, which leads to a decrease in the net throughput of the wireless link. Moreover, the ratio of training symbols to information symbols increases as the bandwidth of the wireless signal, the number of transmitter and receiver antennas, and the power the spectral efficiency of the communication system increase. The problem becomes critical in broadband communication systems and in communication systems that apply multiple antennas at the transmitter and/or at the receiver, where the number of training symbols becomes comparable to the number of information symbols.

The ratio between symbols conducting information and symbols for training increases with the generation of communication systems. The first generation of wireless cellular communication systems (1G) Nordic Mobile Telephone (NMT) is based on analog narrowband Ultra High Frequency (UHF) communication in 25~\!kHz radio channels. The NMT voice channel is transmitted with frequency modulation (FM). NMT supports handover between cells, which requires signaling and channel quality monitoring. NMT signaling uses Fast Frequency Shift Keying (FFSK) digital modulation. Signaling between the base station (BS) and the mobile station occurs over the same RF channel used for audio communication. Consequently, users are disturbed by periodic short noise pulses during communication \cite{HTTP:NMT}. The NMT applies frequency duplex for uplink and downlink communication and frequency division multiplex to allow multiple users to communicate simultaneously in the same cell. The link quality is tested during the call by the BS sending a control signal that is returned by the mobile station. Based on the characteristics of the returned signal, BS estimates the signal-to-noise ratio in the downlink and uplink channels. The amount of signaling was negligible because the signal quality is estimated based on the received signal strength.

Second-generation (2G) \cite{redl_1995} wireless communication systems benefit from the use of wider radio channels and digital technology. The GSM system uses 200~\!kHz, where selective frequency fading can be caused by multipath propagation. To cope with multipath propagation, the training sequence is inserted in the middle of the time slot. In 2G systems, approximately one training symbol is transmitted per four information symbols, while this ratio is increased in the later wireless cellular systems (3G, 4G, and 5G). Since the main goal in designing a wireless communication system is to achieve high system spectral efficiency, the number of training symbols must be reduced, which may lead to erroneous estimation of CSI. However, the static parameters of the wireless channel depend only on the position of the terminal in a static environment, so CSI can be retrieved from the CSIs of previous communications stored in the central or local database~\cite{Pesko_REM_2014}. Information about CSI is retrieved from the database when the communication is initialized, reducing the number of training symbols required to estimate CSI. This type of communication is referred to as environment-aware communication. There is a plethora of wireless technologies with technology-specific quality of extracted CSI. In this context, we will provide an overview of the wireless technology-dependent quality of CSI extraction and explore the potential of using CSI in environment-aware wireless communications.

The paper is organized as follows. This introduction is followed by the definition of channel impulse response (CIR) and channel state information. After that, an overview of the potential of wireless cellular technologies for estimating CSI is given. The next section takes a closer look at CSI estimation in wireless local area networks (WLANs). We then explore the potential of channel estimation in wireless sensor networks (WSNs). We then discuss the impact of channel variations due to either terminal movement or environmental changes. Discussion and conclusions are provided in the last two sections.

\section{Channel state information and channel impulse response}
\label{sec:CSI_CIR}
CSI describes the properties of the communication channel. It describes how radio waves propagate from the transmitter to the receiver. The content of CSI depends on the type of communication system. While channel attenuation is sufficient for narrowband, flat-fading communication channels, channel impulse response (CIR) is desired for broadband communication systems. The CSI can be instantaneous or statistical. The estimation of the CSI can be data aided, where the training symbol is added to the information signals, or blind, where the CSI is estimated from the statistical properties of the transmitted signals. Modern wireless communication systems use data-assisted CSI estimation~\cite{Wiki_CSI} due to its fast and accurate CSI estimation. In addition, wireless channel attenuation, the level of the received signal or Received Signal Strength Indicator (RSSI), and Reference Signal Received Power (RSRP) or signal-to-interference and noise ratio are often applied for coarse characterization of the CSI. The CIR or Power Delay Profile is often applied for the characterization of the frequency-selective wireless channels.

The basic representation of the wireless channel is given by CIR $h(\tau)$. It can be considered as a linear filter~\cite{Molischa_Wireless_2012}. If the interfering objects, the transmitter and the receiver do not change their position or orientation, the wireless channel or its representation of CIR is static. However, in modern wireless communication, the position and orientation of the transmitter and receiver change, and interfering objects such as vehicles, leaves, people, or doors and windows may also change their position or orientation, which may affect the CIR. In this context, the CIR is described by a time-varying function $h(t,\tau)$, where $t$ is the instantaneous (absolute) time and $\tau$ is the propagation delay. The theory of linear time-variant systems can be applied to the theoretical analysis of real wireless channels.

The wireless channel can be represented as an algebraic function in the time domain as CIR $h(t,\tau)$ or in the frequency domain as the channel transfer function $H(t,f)$. The channel transfer function and CIR are related by the Fourier transform. However, the algebraic representation is not suitable to be stored in the database, so the discrete function representation is often used
\[
H(n) = \sum_{k=0}^{N-1} h(k) e^{-j 2 \pi \frac{n}{N} k}
\]
for $n = 0, 1, 2, \cdots , N-1$ and $k = 0, 1, 2, \cdots , N-1$. The discrete representation of CIR is expressed for a given sampling frequency 
\[f_s = \frac{1}{T},\] 
where $T$ is the sampling interval. The bandwidth of the channel transfer function $H$ depends on the sampling frequency $f_s$ and lies in the frequency interval $\left [-f_s/2, f_s/2 \right ]$, while the frequency step $\Delta f$ depends on the number of samples 
\[\Delta f=\frac{f_s}{N}.\]
To provide complete information about the channel transfer function $H(n)$ or the CIR $h(k)$, the information about the sampling interval $T$ or the sampling frequency $f_s$ should be stored in the database. The CSI is often specified as a power delay profile (PDP), which gives the intensity of a signal received over a multipath channel as a function of time delay and can be expressed as CIR $\|h(t, \tau)\|^2$.

In order to introduce environment-aware wireless communications, in the next sections we explore the potential of selected wireless communications to estimate the CIR and accessibility of CIR outside the system.

\section{Wireless cellular technologies}
\label{sec:Cellular_tech}

\begin{figure*}[!h]
\centering
\includegraphics[trim=10 10 10 80, clip, width=1.8\columnwidth]{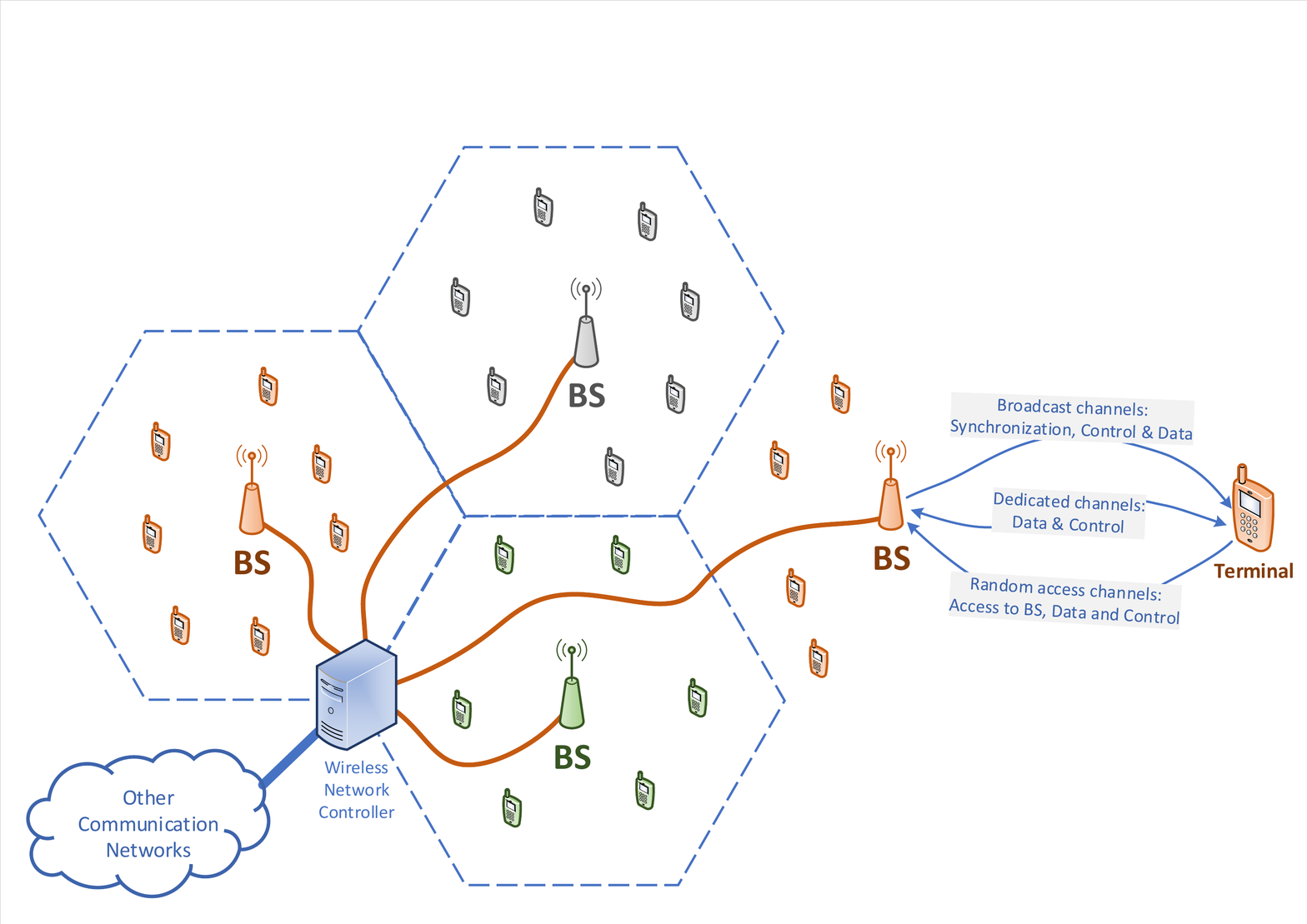}
\caption{The concept of wireless cellular communication and physical radio channels.}
\label{fig:model_of_cell_com_sys}
\end{figure*}

The concept of wireless cellular networks was introduced to achieve higher system capacity with the limited bandwidth allocated to the mobile networks to cope with the congestion in radio networks using a single cell. Today, wireless cellular communication systems provide near-global coverage and are therefore candidates for providing wide-area links for CSI estimation on various carrier frequencies. Since the introduction of cellular networks in the late 1970s, wireless technology has evolved significantly, from the first generation (1G) to the latest (5G). Each generation of wireless cellular communication introduces some new concepts, making it significantly different from the previous one. However, the main concept of wireless cellular networks has not changed significantly. The wireless cellular system consists of a set of base stations (BS) distributed over the area and a set of terminals often referred to as user equipment (UE) or mobile stations (MS). The BSs are interconnected by wired or wireless links in a network. Radio resources in the frequency, time, code, or space domains are allocated by a wireless network to particular communication between BS and the terminal. To successfully connect to the network and establish communication with another terminal device inside or outside the source communication network, BS and the terminal communicate through physical wireless channels, which represent the bearer of one or more logical channels specified and used in the Medium Access Control (MAC) layer. The structure of the physical channel, i.e. modulation and coding scheme, and the amount of the training sequence depend on the information transmitted over the physical channel and the expected channel impairments in the wireless network. The concept of wireless cellular communication is illustrated in Fig.~\ref{fig:model_of_cell_com_sys}.

In any wireless cellular communication system there exist several physical channels. We distinguish between downlink physical channels, i.e. transmission from BS to the terminal, and uplink physical channels, i.e transmission from the terminal to BS. In the downlink, some physical channels are destined for all terminals in the BS range, while other physical channels are intended only for a particular terminal. We call the formal channels broadcast physical channels, while the latter are referred as terminal-dedicated physical channels. Broadcast physical channels include  
 \begin{inparaenum}[(i)]
    \item the synchronization channels,  used for carrier frequency, time, and frame synchronization and transmission power adjustment, 
    \item the control channels used for broadcasting control information to all terminals in the BS range, and  
    \item the data channels used for broadcasting information about the BS functionality as well as data directed to all terminals in the range.
\end {inparaenum} 

The set of terminal-dedicated physical channels includes two types of channels, namely control and data channel. The dedicated channels in the downlink are usually paired with the uplink physical data and control channels.  

In the uplink, there exists also random access channels that are applied to request access to BS, send data, and control to BS based on the ALOHA approach. The classes of physical radio channels between BS and terminal are illustrated in Fig.~\ref{fig:model_of_cell_com_sys}. The arrows illustrate the direction of communication. Each generation of the wireless cellular system uses slightly different terminology for the BS, the terminal, and the physical channel naming, mainly to distinguish the technology from each other. 

In this context, we provide a critical overview of CSI estimation in existing and emerging wireless communication systems that we believe will be in operation in the next few years. A particular focus is on CIR estimation. The overview includes older digital wireless systems such as GSM and UMTS, which are currently considered obsolete and unlikely to be deployed in future wireless systems, as well as wireless systems that are expected to be in operation in the next few decades, namely LTE and 5G.

\subsection{Global System for Mobile Communications (GSM) - the second generation (2G)}
\label{subsec:GSM_tech}
\begin{figure}[!h]
\centering
\includegraphics[trim=0 70 400 300, clip, width=1\columnwidth]{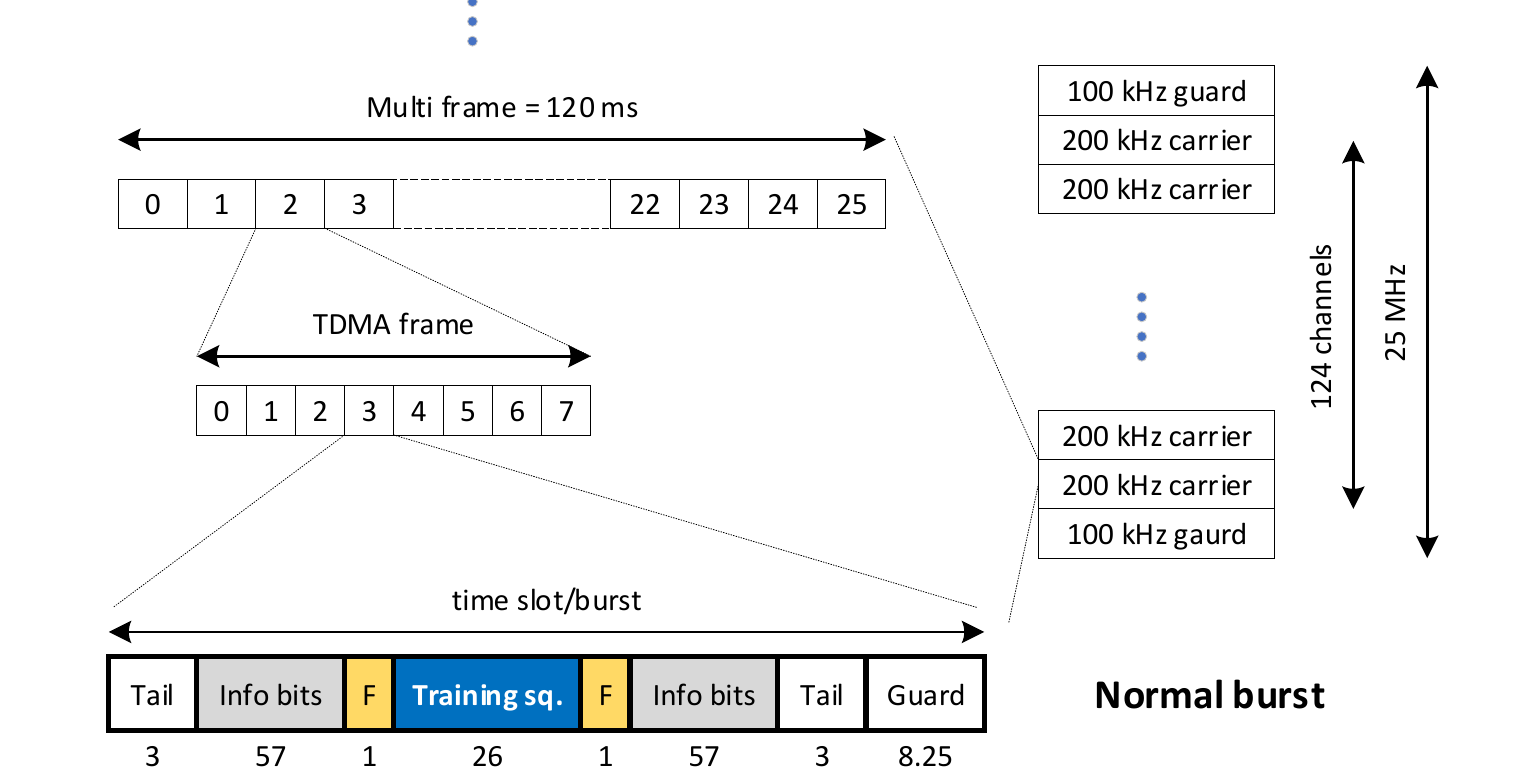}
\caption{The GSM time/frequency multiple access.}
\label{fig:GSM_frame_structure}
\end{figure}

The GSM system is based on the 1G wireless cellular system, which uses frequency division multiple access (FDMA) to separate users. In addition, time division multiple access (TDMA) has been used to take advantage of digital technology, so GSM combines TDMA and FDMA multiple access~\cite{redl_1995}. Initially, the bandwidth of 25~\!MHz is divided into 124 carrier frequencies spaced 200~\!kHz apart. Each carrier frequency is divided into eight time slots, i.e., eight channels, with a duration of 0.577~\!ms (15/26~\!ms). Eight of the time slots are combined to form a TDMA frame, which is 4.615~\!ms (i.e. 120/26~\!ms) long. The frames are organized into so-called multi- and super-frames to ensure overall synchronization. While the time slot is the time allocated to a particular user, the transmission that takes place in a time slot is the GSM burst and carries physical channels. The GSM frame structure is illustrated in Figure~\ref{fig:GSM_frame_structure}.

The GSM standard specifies four types of bursts/physical channels, namely normal bursts (in the uplink and downlink), synchronization bursts in the downlink only, frequency correction bursts in the downlink only, and random access (bursts) in the uplink only. In the GSM system, the terminal is called User Equipment (UE), and BS is a base transceiver station (BTS). Several BS are interconnected and controlled by the Base Station Controller (BSC), whose main tasks are to manage the UE, manage the radio resources, monitor and control the handover between BS, and establish the connection between BS and the Mobile Switching Center (MSC). The assignment of the GSM bursts to the classes of physical radio channels is plotted in Table~\ref{tab:GSM_channels}.

\begin{table}[b!]
\centering
\caption{Types of GSM physical channels, GSM bursts, and their applicability for CIR estimation.}
\label{tab:GSM_channels}
\resizebox{0.85\columnwidth}{!}{%
\begin{tabular}{c|c|c}
\textbf{GSM burst} & \textbf{Usage}     & \textbf{Comments} \\ [4pt] \hline \hline \tstrut
\multirow{4}{*}{normal}   & uplink,     & narrow band 200~\!kHz, \\
                          & downlink,   & 5 tap CIR, \\
                          & data,       & max. delay spread:  \\
                          & control,    & 18.5~\!$\mu$s  \\ [4pt] 
                          \hline \tstrut
\multirow{2}{*}{random access}  & uplink,     &  not frequent \\
                                & random access &  transmission \\ [4pt] \hline \tstrut
\multirow{2}{*}{frequency correction}   & downlink,         & narrow band, \\
                                        & synchronization   & poor autocorrelation \\ [4pt] \hline \tstrut
\multirow{2}{*}{synchronization}    & downlink,         & narrow band, \\
                                    & synchronization  & poor autocorrelation\\ [4pt] \hline
\end{tabular}%
}
\end{table}

The normal burst is applied for communication between BS and UE. The normal burst consists of a 26-bit long training sequence. Eight different training sequences are specified in the GSM standard. The same training sequence is used in each GSM slot transmitted by the same BS while neighboring BSs using the same radio frequency channels use different sequences. The GSM training sequence allows the receiver to estimate the CIR with five complex taps~\cite{redl_1995} separated by a GSM symbol duration of 3.69~\!µs, resulting in an excess delay of almost 18.464~\!µs. 

The purpose of the GSM synchronization burst is transmitted from BS to help the mobile station synchronize with the network. Due to its poor auto- and cross-autocorrelation properties, it is not suitable for estimating CIR. The frequency correction burst consists of 3 tail bits at the beginning and end of the burst, 8.25 bits of guard time, and 142 bits set to zero. The single-tone signals are not suitable for estimating CIR due to their narrow band characteristics. The random access burst is a short burst with a longer guard time that allows the mobile terminal to access the BS. The Random Access Burst is transmitted only when the mobile terminal wants to access the wireless cellular network. This happens from time to time, which limits its use in estimating CIR.

The GSM system can be used to estimate the CIR, but the channel is a narrowband channel with only 200~\!kHz, and the training sequence of a normal burst can only estimate the multipath excess delay of 18.464\~\!µs. If possible, another technology should be applied to estimate the CIR.

\subsection{Universal Mobile Telecommunications System (UMTS) - the third generation (3G)}
\label{subsec:UMTS_tech}
\begin{figure}[!h]
\centering
\includegraphics[trim=10 20 300 250, clip, width=1.0\columnwidth]{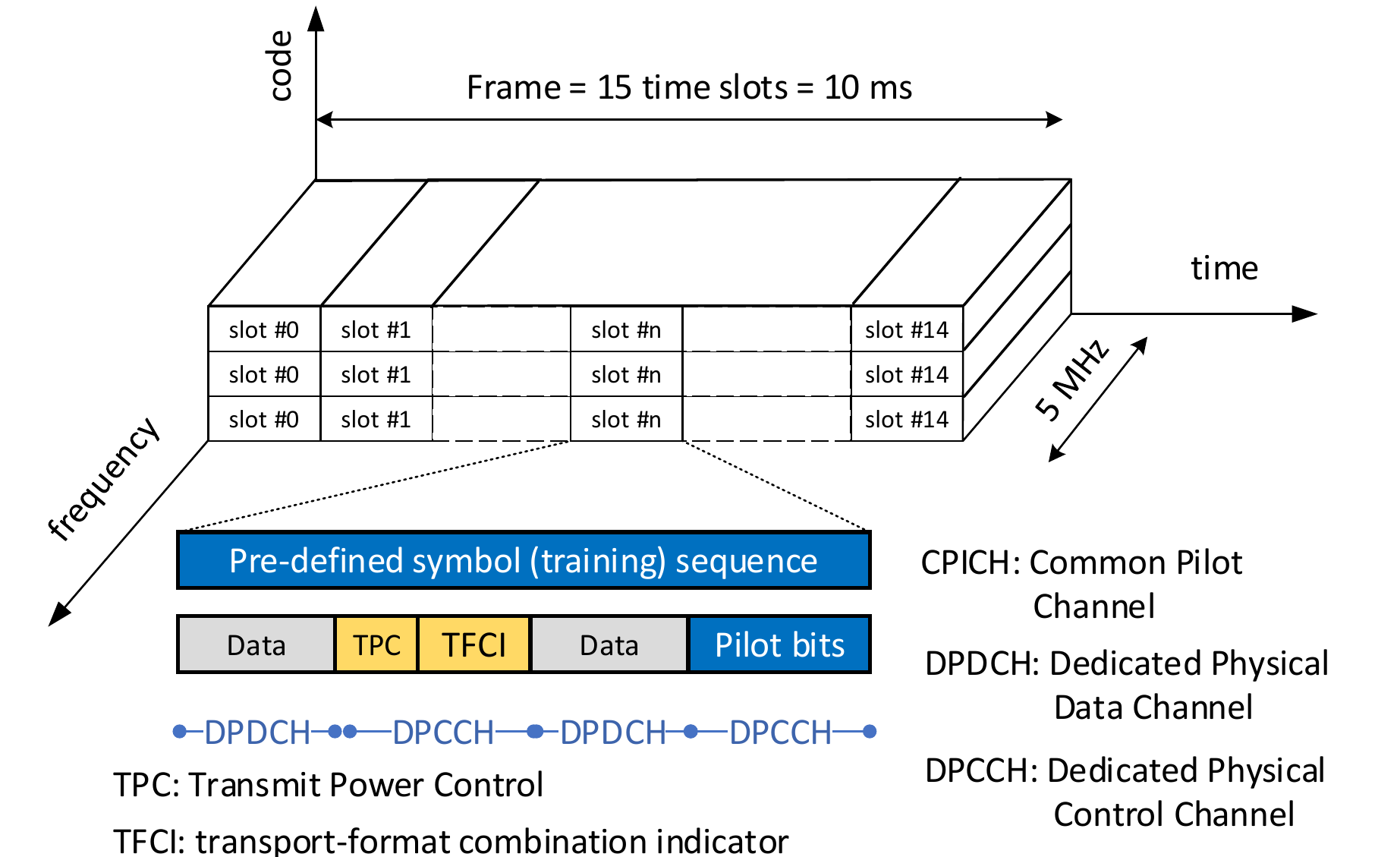}
\caption{The UMTS frame structure.}
\label{fig:UMTS_frame_structure}
\end{figure}

The Universal Mobile Telecommunications System (UMTS) is a third-generation (3G) wireless cellular communications system. The system architecture of the radio part consists of User Equipment (UE), the base station, called Node~\!B, and the Radio Network Controller. The Node~\!B provides the connection between the UE and the telephone network. The Radio Network Controller is responsible for controlling the Node~\!Bs connected to it. UMTS uses wideband code division multiple access techniques in 5~\!MHz frequency channels. It supports frequency division and time division duplex transmission. The Orthogonal Variable Spreading Factor code (OVSF) is used to spread the signal in the 5~\!MHz channel. The OVSF codes have a low correlation with each other as long as there are no delays or echoes in the transmission channel. OVSF-based spreading also allows different spreading factors in different channels. OVSF spreading is also referred to as channelization codes. The channelation code is used to separate the physical and control channels in the uplink, while in the downlink they are primarily used to separate the different users. The channelization codes increase the signal bandwidth. In addition to channelization codes, UMTS also uses scrambling codes, which are used in the uplink to separate terminals/users and in the downlink to separate the transmission of cells. The UMTS frame structure is depicted in~\ref{fig:UMTS_frame_structure}.

Similar to GSM, the physical channels are classified into broadcast, dedicated, and random access channels. There are several UMTS physical broadcast channels \cite{ETSI:TS.25.211}. They are referred to as downlink common physical channels and include \emph{(i) control channels}, namely the primary and secondary common control physical channel (P-CCPCH and S-CCPCH), the acquisition indication channel (AICH), a paging channel (PCH) and physical channels for the common packet channel access procedures; \emph{(ii) data channels}, namely, physical downlink shared channel (PDSCH); and \emph{(iii) synchronization channels}, namely the synchronization channel (SCH) and the common pilot channel (CPICH). The common pilot channel CPICH is the main channel for power and CSI estimation in the downlink. The CPICH is broadcast by the Node~\!Bs at constant power. It can be used for signal quality measurements available as received signal code power (RSCP) and signal-to-noise ratio Ec/No.

UMTS uplink access channels include the Physical Random Access Channel (PRACH) and the Physical Common Packet Channel (PCPCH), which are used to request access to the Node~\!B or to transmit data based on the random multiple access protocol. These channels do not contain sufficient pilot symbols to be applied for the CIR estimation.

In both the uplink and downlink, there are dedicated physical channels for control and data transmission. The Dedicated Physical Control Channel (DPCCH) is the most important channel for estimating CIR. In the uplink, the DPCCH  is a twin channel to the Dedicated Physical Data Channel (DPDCH), and is multiplexed in the I/Q part of the DPDCH, while in downlink the time multiplexing is applied. DPCCH is used to carry control information consisting of known pilot bits to support channel state estimation for coherent detection, transmit power control (TPC) commands, feedback information (FBI), and an optional transport format combining indicator (TFCI). The UMTS physical channels with their relation to the estimation of CIR are listed in Table~\ref{tab:UMTS_channels}.
 
It is foreseen that the 3G communication systems will be replaced in the near future by the 4G and 5G systems, which provide a higher quality of service with less complicated, i.e., cheaper, communication systems for deployment and maintenance, so the mobile operators are shutting down the 3G networks. However, the UMTS communication systems usually use RAKE receivers. The RAKE receiver estimates the CIR, i.e., the phase shift and attenuation of each finger of the RAKE receiver so that the 5~\!MHz radio CSI in UMTS can be estimated. The resolution of the time delay of the channel state is related to the chip duration, i.e., the frequency of the spreading code, which is 3.84~\!Mcps or 0.26~\!µs in the UMTS system, which corresponds to 78~\!m. The maximum delay spread in the rural hilly environments for the UMTS channel is expected to be 20~\!µs~\cite{ETSI:TS.25.943}. The UMTS RAKE receiver should cope with such excess delay, which means the CIR is calculated within the receiver, but unfortunately not available outside the receiver.  

\begin{table*}
\centering
\caption{Types of UMTS physical channels, and their applicability for CIR estimation.}
\label{tab:UMTS_channels}
\resizebox{2.00\columnwidth}{!}{%
\begin{tabular}{c|c|c}
\textbf{Channel type} & \textbf{Channel name} & \textbf{Comments} \\[4pt] \hline \hline \tstrut
\multirow{8}{*}{Broadcast}  &  primary common control physical channel (P-CCPCH)     &  \multirow{2}{*}  {pilot bits for RSCP and Ec/No} \\
                            & secondary common control physical channel (S-CCPCH)   &  \\[4pt]
                            \cdashline{2-3}[1pt/3pt] \tstrut
                            & acquisition indication channel (AICH) & no pilot bits \\[4pt]
                            \cdashline{2-3}[1pt/3pt] \tstrut
                            & paging ch. (PCH) & no pilot bits \\[4pt]
                            \cdashline{2-3}[1pt/3pt] \tstrut
                            & physical channel for the common packet ch. access procedures & not relevant for CIR \\[4pt]
                            \cdashline{2-3}[1pt/3pt] \tstrut 
                            & physical downlink shared channel (PDSCH) & no pilot bits \\[4pt]
                            \cdashline{2-3}[1pt/3pt] \tstrut
                            & synchronization channel (SCH) & relevant for synchronization \\[4pt]
                            \cdashline{2-3}[1pt/3pt] \tstrut
                            & common pilot channel (CPICH) & help in estimation of CSI \\ [4pt]
                          \hline \tstrut
\multirow{3}{*}{Dedicated}  & dedicated physical control channel (DPCCH)    &  contain pilot bits for CSI estimation \\[4pt]
                            \cdashline{2-3}[1pt/3pt] \tstrut
                            & dedicated physical data channel (DPDCH)       &  no pilot bits \\[4pt]
                            \cdashline{2-3}[1pt/3pt] \tstrut 
                            &                                               & in uplink PDCCH and PDDCH are I/Q multiplexed \\
                            &                                               & in downlink PDCCH and PDDCH are time multiplexed \\ [4pt] \hline \tstrut
\multirow{2}{*}{Random access}  & physical random access channel (PRACH)    & only for uplink power adjustment \\[4pt]
                            \cdashline{2-3}[1pt/3pt] \tstrut
                                & physical common packet channel (PCPCH)    & random access in uplink, not relevant \\[4pt] \hline 
\end{tabular}%
}
\end{table*}

\subsection{Long Term Evolution (LTE) - the fourth generation (4G)}
\label{subsec:LTE_tech}

\begin{figure}[!h]
\centering
\includegraphics[trim=55 55 160 10, clip, width=1.0\columnwidth]{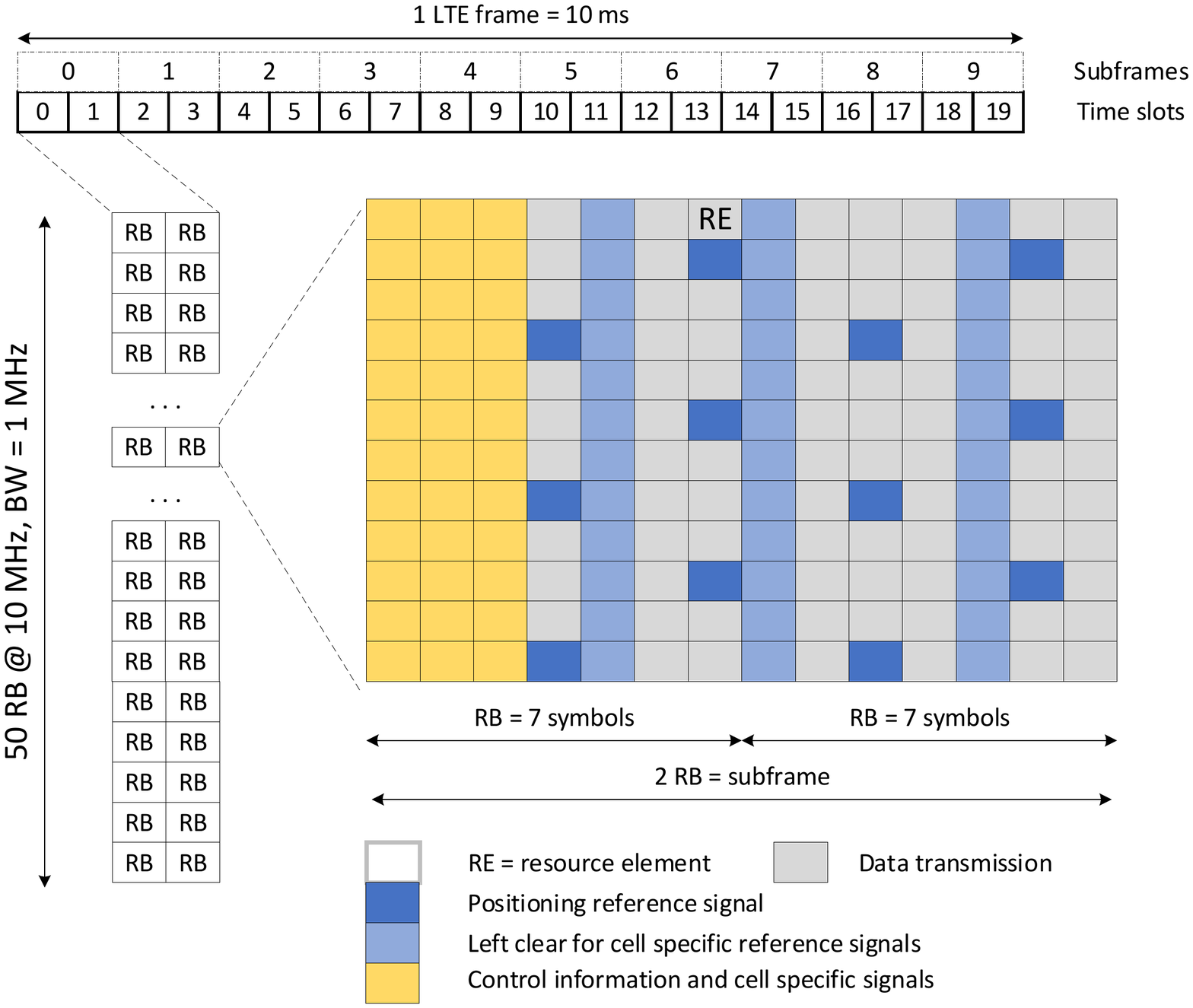}
\caption{The LTE frame structure.}
\label{fig:LTE_frame_structure}
\end{figure}

Long-Term Evolution (LTE) is the evolution of UMTS towards an all-IP broadband network. The radio access part of LTE, Evolved Universal Terrestrial Radio Access (E-UTRA), provides a framework for increasing data rates and overall system capacity, reducing latency, and improving spectral efficiency and performance at cell edges \cite{Johnson_Long_2012}. It is designed to support IP connections. The BS in LTE is called eNode~\!B. E-UTRA applies orthogonal frequency-division multiplexing (OFDM) in the downlink and both OFDMA and a precoded version of OFDM called Single-Carrier Frequency-Division Multiple Access (SC-FDMA) in the uplink. In both OFDM and SC-FDMA transmission modes, a cyclic prefix is appended to the transmitted symbols to handle multipath propagation. LTE supports both frequency and time division duplex transmission. The basic communication element in LTE is a time-frequency block called a resource block (RB). The duration of the resource block is 0.5~\!ms. It consists of seven time symbols with a bandwidth of 180~\!kHz. Each symbol contains 12 subcarriers with a bandwidth of 15~\!kHz. The resource element (RE) represents a time-frequency unit with the duration of one symbol and the bandwidth of one subcarrier. One symbol occupying all 12 subcarriers is called RB~\!column in this text and thus consists of 12 REs. The resource blocks are distributed over several frequency bands with a bandwidth of 1.4~\!MHz, 3~\!MHz, 5~\!MHz, 10~\!MHz, 15~\!MHz and 20~\!MHz. In the time domain, LTE transmission is structured in radio frames. Each of these radio frames is 10~\!ms long and consists of 10 subframes of 1~\!ms each, i.e., two RBs. The frame structure of the LTE system is drawn in Figure~\ref{fig:LTE_frame_structure}. 

In LTE and LTE advanced, there are multiple physical downlink and uplink channels. The physical channels, which carry data are called shared channels, while the channels that relay the control information are named control channels. The physical shared channels can also carry control information. In addition, there are channels to support broadcast and multicast data, namely the physical broadcast channel and the physical multicast channel, and special physical channels to transport important control information, namely the physical control format indicator channel and the physical hybrid automatic repeat request indicator channel in the downlink, and the physical random access channel in the uplink.

The training signals in LTE are not part of the physical channels but are specified as reference signals (RS) that are transmitted within selected resource elements. In the downlink, there are several reference signals associated with different downlink physical channels. The signals broadcast to all users are synchronization signals, cell-specific reference signals, mobile broadcast multimedia service single-frequency network reference signals, positioning reference signals, and CSI reference signals. The user equipment-specific or demodulation reference signal is dedicated to a single user. In the uplink, only three reference signals exist, namely, the sounding reference signal, the demodulation reference signal for the physical uplink shared channel (PUSCH),  and the demodulation reference signal for the physical uplink control channel (PUCCH). The LTE reference signals are listed in Table~\ref{tab:LTE_signals}. 

Synchronization signals (SS) are broadcast signals used to synchronize frames, subframes, slots, and symbols in the time domain and to determine the center of the channel bandwidth. SS is transmitted twice in each frame. There is a primary SS, which is transmitted on the 62 central subcarriers at the last symbol in time slots 0 and 10, and a secondary SS, which is transmitted on the 62 central subcarriers at the second to last symbol in time slots 0 and 10. The SS can be applied for coarse CIR estimation in the center of the transmission band, but the training signals occupy only narrow bandwidth of 930~\!kHz.

Cell-specific reference signals are broadcast over the entire cell. They are used to estimate the Reference Signal Received Power (RSRP), Reference Signal Received Quality (RSRQ), and Channel Quality Indicator (CQI). It corresponds to the Common Pilot (CPICH) channel in UMTS. The cell-specific reference signals are transmitted in subframes that support the transmission of the physical downlink shared channel. They can be applied for CSI estimation in bandwidth or a single resource block of 180~\!kHz.

The multimedia broadcast multicast (MBM) transmitted in subframes single frequency network (SFN) (MBMSFN) reference signals are transmitted in the resource blocks allocated for the MBM service, which is today rarely used. Similarly, as for cell-specific reference signals CIR estimation is obtained for a single RB with a bandwidth of 180~\!kHz.

LTE provides indoor positioning capabilities. To support this functionality the positioning reference signals (PRS) are broadcast in the predefined positioning subframes. The PRS signals are transmitted over the entire frequency BS range and are therefore useful for estimating CIR. In addition, no data is transmitted in the resource blocks selected to transmit the PRS signal from any eNode~\!Bs in the network.

The user equipment (UE) specific reference signals also referred to as demodulation reference signals are transmitted in the resource blocks allocated for the PDSCH. They co-exist with the cell-specific reference signals. They were introduced primarily to support signal demodulation, beamforming, and MIMO transmission. They are transmitted while the eNode~\!B is connected to the UE. 

Resource elements allocated for data transmission often experience different interference from resource elements allocated for cell-specific reference signals. This difference is in particular observed for the unloaded networks. In order to improve this and enable MIMO and multiuser MIMO transmission, the CSI reference signals are specified in the LTE standard. They are transmitted in the RB assigned for PDSCH. In order not to interfere with data transmission the CSI reference signals are transmitted in one subframe per 5, 10, 20, 30, or 80 frames. The CSI reference signals are only transmitted when communication is active, which limits the use of the CSI signals.

Sounding Reference Signal (SRS) is a reference signal sent from the UE to the eNode~\!B to estimate the quality of the uplink channel over a wider bandwidth. The eNode~\!B can use this information for uplink scheduling of data bits to specific subcarriers. The eNode~\!B can also use SRS for timing estimation in the uplink, as part of timing alignment, especially in situations where there are no PUSCH/PUCCH transmissions in the uplink for an extended period of time.

The demodulation reference signal associated with the Physical Uplink Shared Channel (PUSCH) is applied to support the demodulation of the uplink transmission in the PUSCH. They occupy 12 subcarriers in the middle of the resource block so that it can be used for channel estimation for channels with a bandwidth of 180~\!kHz and when the UE is connected to the eNode~\!B.   

There are several formats of Physical Uplink Control Channel (PUCCH) reference signals, namely 1, 1a, 1b, 2, 2a, 2b, and, 3. Three columns in the time-frequency block  are dedicated to the reference signal in PUCCH formats 1, 1a, and 1b, while there are only two for formats 2a, 2b, and 3. 

LTE and LTE advanced technologies offer tremendous potential for monitoring \textit{CSI} in multiple frequency bands. However, with current specifications, the information from CSI cannot be retrieved by the system through standardized application programming interfaces. However, the basic concepts of LTE and LTE Advanced are adopted by the next generation of mobile radio, "New Radio".

\begin{table*} [!t]
\centering
\caption{Types of LTE reference signals and their applicability for CSI estimation.}
\label{tab:LTE_signals}
\resizebox{2.00\columnwidth}{!}{%
\begin{tabular}{c|c|c}
\textbf{Signal type} & \textbf{LTE reference signals (RS)} & \textbf{Comments} \\ [4pt] \hline \hline \tstrut
\multirow{8}{*}{Broadcast}  
    & synchronization signals (SS)    &  {coarse CSI estimation, BW = 930~\!kHz} \\ [4pt]
    \cdashline{2-3}[1pt/3pt] \tstrut
    & cell-specific RS   &  {transmitted in subframes allocated for PDSCH, BW = 180~\!kHz} \\ [4pt]
    \cdashline{2-3}[1pt/3pt] \tstrut
    & multimedia broadcast multicast (MBM) & transmitted in subframes\\ [4pt]
    & single frequency network (SFN) RS &  allocated for MBM service, BW = 180~\!kHz\\ [4pt]
    \cdashline{2-3}[1pt/3pt] \tstrut
    & positioning RS (PRS) & transmitted in the whole LTE band \\ [4pt]  
    \hline \tstrut
\multirow{3}{*}{Dedicated - downlink} 
    & user equipment  specific (demodulation) RS  & transmitted when UE is connected \\ [4pt]
    \cdashline{2-3}[1pt/3pt]
    & channel state information (CSI) RS       &  transmitted when PDSCH is transmitted \\  [4pt]
    \hline \tstrut
    \multirow{4}{*}{Dedicated - uplink}  
    & sounding RS   & channel quality in uplink \\ [4pt]
    \cdashline{2-3}[1pt/3pt] \tstrut
    & demodulation RS with PUSCH             & when UE is connected to eNode~\!B \\ [4pt]
    \cdashline{2-3}[1pt/3pt] \tstrut
    & demodulation RS with PUCCH             &  when control information is transmitted \\  [4pt]
    \hline 
\end{tabular}%
}
\end{table*}

\subsection{New Radio (NR) - the fifth generation (5G)}

5G wireless cellular technology was developed to meet the requirements for higher data rates and lower latency compared to existing 4G technology~\cite{Johnson_5G_2019}. To achieve this, a scalable and flexible air interface has been developed under 3GPP. Also, the new spectrum is allocated in cm and mm frequency bands in addition to the frequency bands below 6 GHz. The 5G specifications also include support for massive multi-antenna and multi-beam antenna systems. The radio access part of 5G systems is referred to as new radio (NR) and specifies the time and frequency division duplex transmission. A 5G base station is referred to as a gNode~\!B (next generation Node~\!B) to distinguish it from a 4G base station (evolved Node~\!B, eNode~\!B).

NR uses OFDMA similar to LTE, but supports many OFDM subcarrier spacings, namely 15~\!kHz, 30~\!kHz, 60~\!kHz, 120~\!kHz, and 240~\!kHz. Higher subcarrier bandwidth results in shorter symbol duration and consequently lower latency in information transmission. The following symbol durations correspond to the specified subcarrier spacings 66.67~\!µs, 33.33~\!µs, 16.67~\!µs, 8.33~\!µs, and 4.17~\!µs. The transmissions from NR are organized into time frames of 10 ms, consisting of 10 subframes of 1 ms. The subframes contain a variable number of time slots. The number of time slots per subframe depends on the cyclic prefix (normal or extended) and the spacing of the subcarriers. A slot consists of 14 OFDM symbols for the normal cyclic prefix and 12 for the extended cyclic prefix. The resource block in NR is specified differently than in LTE, namely the resource block consists of 12 subcarriers and only one symbol, which means that the RB does not contain a timing component. The frame structure of NR is shown in Figure~\ref{fig:}.

The NR specifies a set of physical channels for transmission in the downlink (Physical Broadcast Channel (PBCH), Physical Downlink Shared Channel (PDSCH), Physical Downlink Control Channel (PDSCH)) and in the uplink (Physical Uplink Access Channel, Physical Uplink Shared Channel (PUSCH), Physical Uplink Control Channel (PUCCH)). The NR also specifies several reference signals used by the physical layer for synchronization, channel quality measurement, and channel estimation. Some of them are closely related to the physical channels. The physical signals can be classified as broadcast signals (synchronization signals (SS), demodulation reference signals for the broadcast physical channel, CSI reference signals, phase tracking reference signals), demodulation reference signals associated with physical channels, namely PUSCH, PUCCH, PDSCH and PDCCH, and uplink system reference signals, namely sounding reference signal and uplink phase tracking reference signals.

\begin{figure}[!h]
\centering
\includegraphics[trim=10 25 270 390, clip, width=1.0\columnwidth]{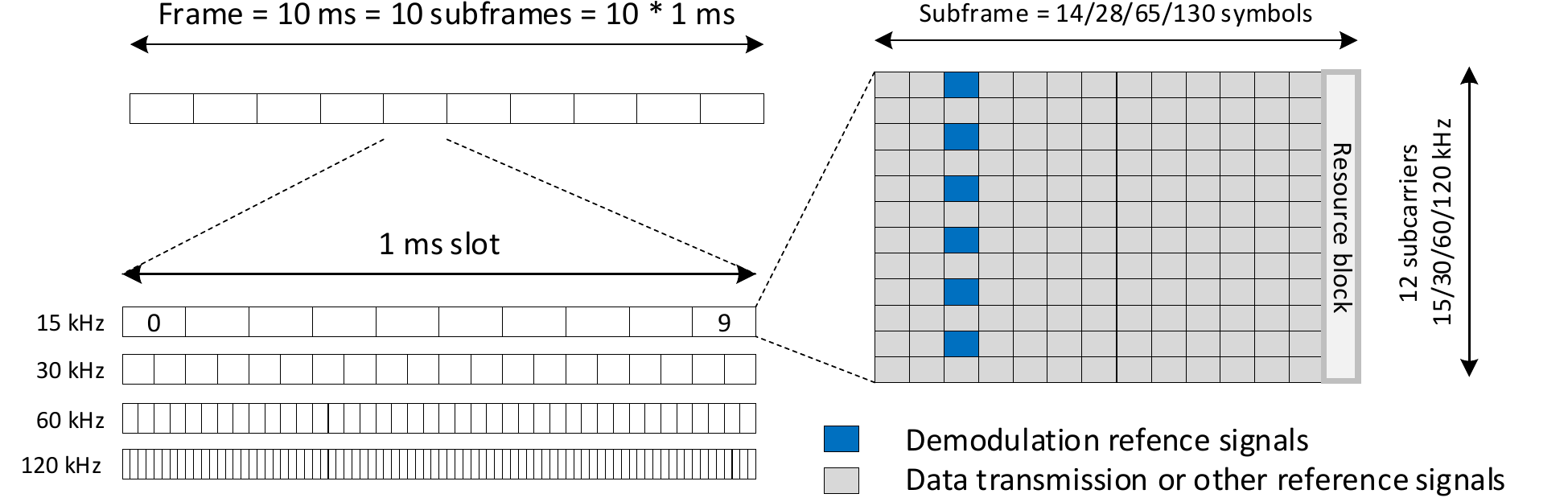}
\caption{The NR frame.}
\label{fig:NR_frame_structure}
\end{figure}

There are two synchronization signals, a primary and a secondary. The primary synchronization signal occupies 127 subcarriers in the first symbol of the frame, while the secondary synchronization signal is transmitted in 127 subcarriers in the third symbol. The 2nd and 4th symbols of the frame and the unoccupied subcarriers in the 3rd symbol are assigned to the physical broadcast channel. BPSK modulation is used for the synchronization symbols. In addition to synchronizing the UE with the base station, PSS and SSS are also used to identify the physical cell identity (PCI), to perform RSRP, RSRQ, and SNIR measurements, and as an additional demodulation reference signal for the PBCH. The bandwidth occupied for the SS depends on the subcarrier spacing and therefore varies between 3.6~\!MHz and 57.6. The SS can be applied for an initial CSI estimate in NR.

The demodulation reference signal (DMRS) for PBCH is inserted into the resource blocks to support the decoding of the information transmitted in the PBCH. The PBCH is transmitted in the 48 RBs in the synchronization signal/PBCH block. 3 DMRS signals are inserted in each resource block, so there are 3*48 = 144 reference signals. The DMRS symbols are generated with a pseudo-noise generator whose seed depends on PCI. The DMRS are QPSK modulated and occupy the bandwidth between 3.6~\!MHz and 57.6~\!MHz 

The CSI reference signal (CSI-RS) is not a new concept in NR, it has already been used in LTE but is more flexible in NR. The transmission of CSI-RS is limited to the entire bandwidth of the NR, but not only to the transmissions in the part of the spectrum where the data is transmitted. The gNode~\!B is now aware of the CSI for the complete NR bandwidth. In the frequency domain, the CSI-RS can be transmitted in all resource blocks or every two resource blocks. In the time domain, CSI-RS transmission can be periodic, aperiodic, and semipersistent. The periodic CSI-RS can be transmitted in the Nth time slot, where N is between 4 and 640. Semipersistent transmission is similar to a periodic transmission, but transmission can be temporarily turned off. In aperiodic mode, transmission is on request from the upper layers. The CSI-RS transmission can be non-zero power or zero power. Transmission with zero power is announced by gNode~\!B to UE, i.e., no signal is transmitted at the specified RB. Consequently, UE can estimate the noise and interference power. The non-zero transmission is mainly used to estimate the channel characteristics. The NR supports the CSI estimation in the whole NR band, even if there is no data transmission.

A tracking reference signal (TRS) is a downlink synchronization signal used by the UE to track time and frequency variations of the carrier frequency and frame timing of BS. The synchronization signal allows course tracking and synchronization, while the tracking reference signal allows high-precision synchronization and tracking. The tracking reference signal is a CSI-RS with special parameters. The assignment of the resource block to TRS RS depends on the configuration of TRS.

The phase tracking reference signal is inserted into RB to cope with phase noise, which is an important degradation at higher frequencies. It can be applied for CSI estimation only at higher frequency bands.

Demodulation Reference Signal (DMRS) for the physical downlink control channel is inserted into the physical downlink control channel to estimate the propagation channel characteristics of the PDCCH. The DMRS occupies 25\% of the resource blocks allocated to the PDCCH, i.e., every fourth subcarrier in the symbol is allocated to the DMRS. The allocation of the DMRS is fixed and does not depend on other network planning parameters.

DMRS for the physical downlink shared channel is always associated with PDSCH. They are used to assist in the demodulation of PDSCH symbols. The resource elements assigned to DMRS are flexible and can be controlled by a set of parameters, which control the number of DMRS in a frequency and time domain. The parameters are determined based on the frequency and time of the channel. The CSI estimation of DMRS is limited to the RB bandwidth

The NR uses multiple uplink reference signals to assist the BS in signal demodulation, channel estimation and supporting measurements. The demodulation reference signals for PUCCH and PUSCH aid in signal demodulation. The sounding reference signal (SRS) is aimed to assist in channel estimation and measurements. The phase tracking reference signals (PT-RS) are used for phase noise compensation.

Transmission of the DMRS associated with PUCCH enables coherent detection. Several formats of PUCCH are specified, and DMRS transmission depends on the PUCCH format. DMRS can be transmitted in the whole resource block, but every second symbol is allocated to PUCCH, or every symbol but only on every third subcarrier, or only on the specific symbols assigned to PUCCH. The CSI can be estimated only during active transmission.

The PUSCH is always combined with the DMRS. The DMRS allocation depends on the allocation of the PUSCH. The DMRS can be transmitted in every second sub-carrier frequency in each time symbol or at the beginning or middle of the time slot. Additional DMRS can be added for fast channel variation in time or frequency.

The sounding reference signal (SRS) is used to measure the characteristics of the uplink channel. It is transmitted on the BS request. The SRS can occupy 1, 2, or four symbols in the time domain and can be transmitted in up to 272 resource blocks. Within the resource block, only a part of the subcarriers can be allocated for the SRS, namely  the second or the fourth subcarrier. 

The phase tracking reference signal (PTRS) is applied to compensate frequency and phase offset of the UE. It is transmitted in RB allocated for the PUSCH. The phase noise has no significant effect at carrier frequencies below 6~\!GHz, but at higher frequencies, the frequency shift must be compensated. The PTRS is often applied to reduce the amount of DMRS transmitted with PUSCH. The PTRS is transmitted during communication. 

\begin{table*} [!t]
\centering
\caption{Types of NR reference signals and their applicability for CSI estimation.}
\label{tab:NR_signals}
\resizebox{2.00\columnwidth}{!}{%
\begin{tabular}{c|c|c}
\textbf{Signal type} & \textbf{NR reference signals (RS)} & \textbf{Comments} \\ [4pt] \hline \hline \tstrut
\multirow{8}{*}{Broadcast}  
        & synchronization signals (SS)    &  {coarse CSI estimation, BW = 3.6 to 57.6~\!MHz} \\ [4pt]
    \cdashline{2-3}[1pt/3pt] \tstrut
        & demodulation RS for physical broadcast channel (PBCH) &  {with PBCH, coarse CSI estimation, BW = 3.6 to 57.6~\!MHz} \\ [4pt]
    \cdashline{2-3}[1pt/3pt]        
        & channel state information (CSI) RS &  {complete NR frequency band} \\ [4pt]
    \cdashline{2-3}[1pt/3pt]
        & tracking RS&  {BW equal to RS bandwidth} \\ [4pt]
    \cdashline{2-3}[1pt/3pt]
        & phase tracking RS&  {BW equal to RS bandwidth, used only above 6~\!GHz} \\ [4pt]  
    \hline \tstrut
\multirow{7}{*}{Demodulation RS} 
        & at the physical uplink shared channel (PUSCH)  &  UE is connected to BS,  BW equal to RS bandwidth \\[4pt]
    \cdashline{2-3}[1pt/3pt] \tstrut
        & at the physical uplink control channel (PUCCH) &   UE is connected to BS,  BW equal to RS bandwidth \\[4pt]
    \cdashline{2-3}[1pt/3pt] \tstrut
        & at physical downlink control channel (PDCCH) &  UE is connected to BS,  BW equal to RS bandwidth \\[4pt]
    \cdashline{2-3}[1pt/3pt] \tstrut
        & at physical downlink shared channel (PDSCH) &  UE is connected to BS,  BW equal to RS bandwidth \\[4pt]
    \hline \tstrut
\multirow{3}{*}{Uplink/system}  
        & sounding RS   &  channel quality in uplink  \\[4pt]
    \cdashline{2-3}[1pt/3pt] \tstrut
        & phase tracking (PT) RS  & UE is connected to BS \\[4pt]
    \hline 
\end{tabular}%
}
\end{table*}

The 5G standard brings an important innovation to CSI estimation, namely the ability to estimate CIR in the time/frequency blocks where no data is being transmitted. This feature opens up the possibility of using the 5G system to obtain global knowledge about radio channels and spectrum occupancy.

\section{Wireless local area networks (WLANs)}
\label{sec:LAN_tech}
\begin{figure}[!h]
\centering
\includegraphics[trim=10 20 200 450, clip, width=1.0\columnwidth]{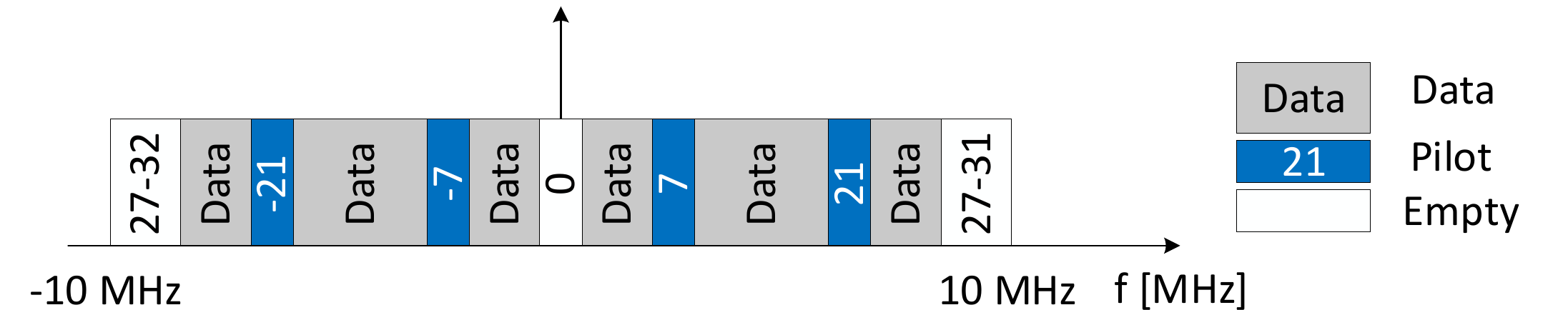}
\caption{The Wi-Fi OFDM burst.}
\label{fig:wifi_OFDM_burst}
\end{figure}

Wireless Fidelity (Wi-Fi) is a low-cost wireless access technology that allows many electronic devices to connect to the Internet. According to the Wi-Fi Alliance, Wi-Fi devices are any Wireless Local Area Network (WLAN) product based on the Institute of Electrical and Electronics Engineers (IEEE) 802.11 standards~\cite{Gast_WiFI_2005}. The IEEE 802 standard refers to a family of IEEE standards that address Local Area and Metropolitan Area Networks. The IEEE ~802.11 family of standards is a subset of IEEE~802 standards that deal with Medium Access Control (MAC) and Physical Layer (PHY) specifications. The 802.11 specifies the number of over-the-air modulation schemes that use the same basic protocol. The first standard in this series is IEEE 802.11.-1997. It specified three alternative physical layer technologies: diffuse infrared, which supports 1~\!Mbit/s, frequency hopping spread spectrum, which supports 1~\!Mbit/s or 2~\!Mbit/s and direct sequence spread spectrum, which supports 1~\!Mbit/s or 2~\!Mbit/s in the 2.4~\!GHz ISM frequency band. Offering too many technology options prevented the widespread adoption of this technology. This was improved in the next version of the standard, IEEE 802.11b, which offers 11~\!Mbit/s peak data rates using 22~\!MHz channels in the 2.4~\!GHz frequency band with the same media access technologies. The standard uses complementary code keying (CCK) modulation, a direct spread spectrum modulation that encodes 4 or 8 bits in 8~\!QPSK symbols. The next version of the standard, IEEE 802.11a, allows the use of the 5~\!GHz frequency band to avoid interference with other systems in the 2.4~\!GHz band ISM, and introduces OFDM modulation, which allows flexible use of the 20~\!MHz frequency band allocated per channel. The OFDM uses 64~\!IFFT. The subcarrier bandwidth is 315.2~\!kHz. The lowest 6 and highest 6 subcarriers are not in use but provide a frequency guard interval. The remaining 52 subcarriers are used for data transmission (48 subcarriers) and four of them, namely -21, -7, 7, and 21, are applied for channel estimation. The symbol duration is 3.2~\!µs, but 0.8~\!µs guard interval or cyclic prefix is added to handle multipath propagation. This results in a symbol duration of 4.0~\!µs. The IEEE 802.11g standard also extends the use of OFDM to the 2.4~\!GHz ISM band. The OFDM burst with guard bands is illustrated in Figure~\ref{fig:wifi_OFDM_burst}.

The 802.11n amendment includes many improvements that increase WLAN range, reliability, and throughput, including new modulation schemes, multiple antennas at the transmitter and receiver, and transmission in the 40~\!MHz bandwidth. The data rate is approaching 600~\!Mbps. The standard supports the 2.4~\!GHz and 5~\!GHz frequency bands. Other enhancements to the IEEE 802.11ac standard include the use of 80-and~\! 160~\!MHz bands support more antennas at the transmitter and receiver, and specification of modulation schemes with high bandwidth efficiency such as 256~QAM. Pilots are transmitted in additional subchannels, e.g., at a bandwidth of 40~\!MHz, the following subcarriers are occupied by pilots: -53, -25, -11, 11, 25, 53.

\begin{figure}[!h]
\centering
\includegraphics[trim=0 20 250 430, clip, width=1.0\columnwidth]{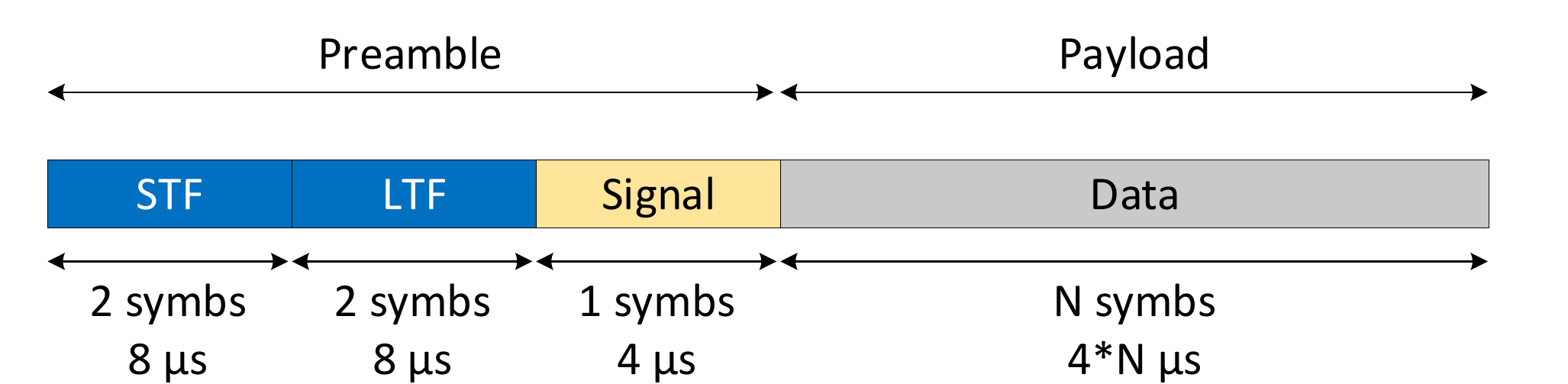}
\caption{IEEE 802.11g time frame.}
\label{fig:Wifi_time_frame}
\end{figure}

In the time domain, transmissions are organized in physical frames. The physical frame of the IEEE standard consists of a preamble and a payload. The preamble consists of a short training field (STF) with two symbols. The STF occupies 25\% of the subcarriers. It is used for the initial time and frequency synchronization. The next part of the preamble is the long training field (LTF), which consists of 2 symbols. The LTF occupies all 52 subcarriers and is used for fine time and frequency synchronization and CIR estimation. This is followed by the signal part, which consists of a symbol containing information about the payload, namely the data rate, i.e. the coding, the length of the payload, the parity information and information about the tail bits. In the standards that support higher transmission rates, the additional information is transmitted in the preamble, which increases the length of the preamble. The duration of the symbol in the preamble varies, from 192 µs for CCK transmission to 20 µs for OFDM transmission according to the IEEE 802.11g standard. The physical layer frame format for IEEE 802.11g is shown in Figure~\ref{fig:Wifi_time_frame}.

The latest Wi-Fi standards support broad bandwidth and allocation of training sig that can be used for CIR estimation. Compared to the cellular system, the Wi-Fi chipset produces allows access to some radio parameters of the technology through well-specified APIs, making this wireless technology a good choice for experiments with CSI estimation at different frequency bands.

\section{Wireless sensor networks (WSNs)}
\label{sec:WSNs_tech}
Recently, wireless sensor networks become very popular to collect information for a huge number of sensors. Sensors spread over a wide area make the technology very convenient for channel estimation in spite of the wireless sensor technologies usually do not support broadband communications occupying a wide frequency band. Among wireless sensor technologies, those based on IEEE 802.15.4 and IEEE 802.15.4a standards are often in use.

\subsection{IEEE 802.15.4 standard}
The IEEE 802.15.4 standard~\cite{Kurunathan_IEEE_2018} specifies the physical layer and media access control for low-rate wireless Personal Area Networks (PANs). The MAC and PHY layers are complemented by various technologies such as ZigBee or RPL (Routing Protocol for Low-Power and Lossy Networks). The devices, which comply with the IEEE 801.15.4 standard, operate in several frequency bands: ISM 2.4~\!GHz band with sixteen 5~\!MHz channels, 915~\!MHz frequency band with ten 2~\!MHz channels, and 868~\!MHz frequency band with only one 0.6~\!MHz channel. The packet data unit (PDU) of the physical layer contains a synchronization header, i.e. a preamble and start of packet delimiter, a PHY header to specify the packet length, and the payload data. The 32-bit preamble is designed for the acquisition of symbol and chip timing and may be used for coarse frequency adjustment. The physical layer supports two methods for estimating channel quality, namely energy detection and link quality indication. Energy detection is an estimate of the received signal power that is primarily used for channel assessment. Link quality indication is measured for each received packet, taking into account the detected packet energy and the signal-to-noise ratio. The standard does not provide sufficient means for CSI estimation.

The IEEE 802.15.4-2011 standard has evolved to the IEEE 802.15.4e standard to meet the QoS requirements of industrial communications. The new MAC behaviors are significantly different from those considered in IEEE 802.15.4-2011. It introduces Time Slotted Channel Hopping (TSCH), which uses fixed-size TDMA slots and multi-channel hopping to provide the ability to monitor CSI across the 2.4~GHz ISM frequency band by sending a continuous wave of constant frequency signal as a payload on different frequency channels.

\subsection{IEEE 802.15.4a standard}
The original 802.15.4 standard cannot meet the requirements of the various applications intended for wireless PANs, especially in terms of transmission rate and precise localization. Therefore, the standard has been extended to include ultra-wideband (UWB) technology, which operates in the unlicensed UWB spectrum~\cite{Karapistoli_Overview_2010}. In addition, the standard specifies a wireless technology to support long-range and reliable communications, but this specification has not yet found many implementations in silicon. Therefore, we will focus on the UWB part of the standard, which is currently the only on-chip technology that provides a documented API to obtain the CSI in the wide frequency band.

Three ISM UWB frequency bands are available, namely the sub-gigahertz band (0.250 to 0.750~\!GHz), the low band (3.244 to 4.742~\!GHz), and the high band (5.944 to 10.234~\!GHz). Within these bands, 16 channels are specified to support different data rates, namely 110 kbits/s, 851 kbits/s, 6.81 Mbits/s, and 27.24 Mbits/s. Channel 0 is allocated in the sub-gigahertz band, channels {1-4} in the low band, and the remaining channels in the high band. A compliant device must operate in at least one of the mandatory channels {0, 3, and 9}. The channel bandwidth is approximately 500 MHz. The range of UWB devices is between 10 and 100 m and the channel access method is CSMA/CA or Aloha 

\begin{figure}[!h]
\centering
\includegraphics[trim=0 20 0 390, clip, width=1.0\columnwidth]{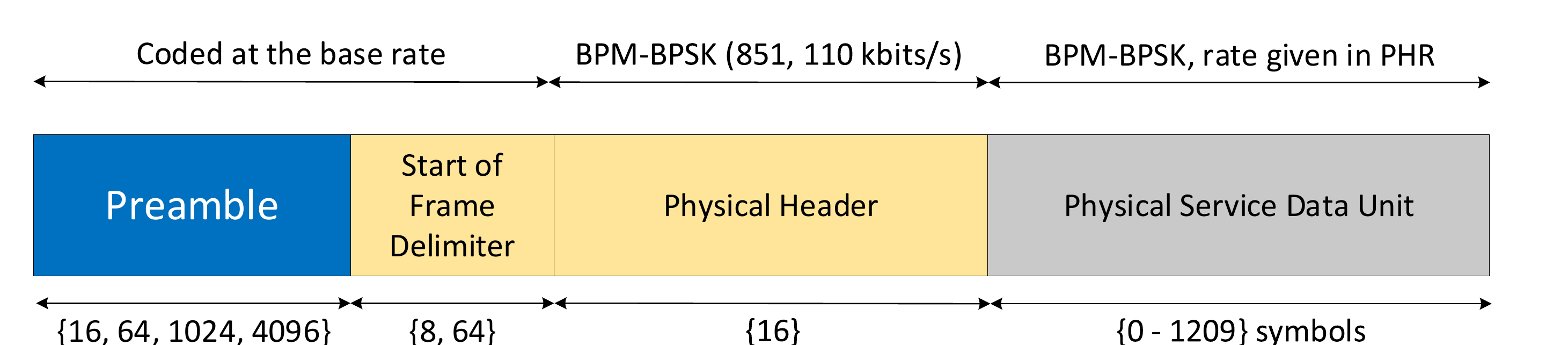}
\caption{IEEE 802.15.4a packet format.}
\label{fig:UWB_packet}
\end{figure}

The packet format of the IEEE 802.15.4a is shown in Figure~\ref{fig:UWB_packet}. The Physical Layer Protocol Data Unit (PPDU) or packet contains a preamble, the Start of Frame Delimiter (SFD), the Physical Header (PHR), and the Physical Service Data Unit (PSDU). The preamble is used for synchronization, frame timing, and CSI estimation. The preamble is transmitted at a nominal data rate of 110 kbits/s. The length of the preamble varies from 16 to 4096. The PHR transmits information for decoding the Physical Service Data Unit (PSDU). The length of the PSDU varies from 0 to 1209 symbols. Several data rates are supported, namely 0.11, 0.85, 6.81, or 27.24~\!Mb/s.

\begin{figure}[!h]
\centering
\includegraphics[trim=0 10 380 380, clip, width=1.0\columnwidth]{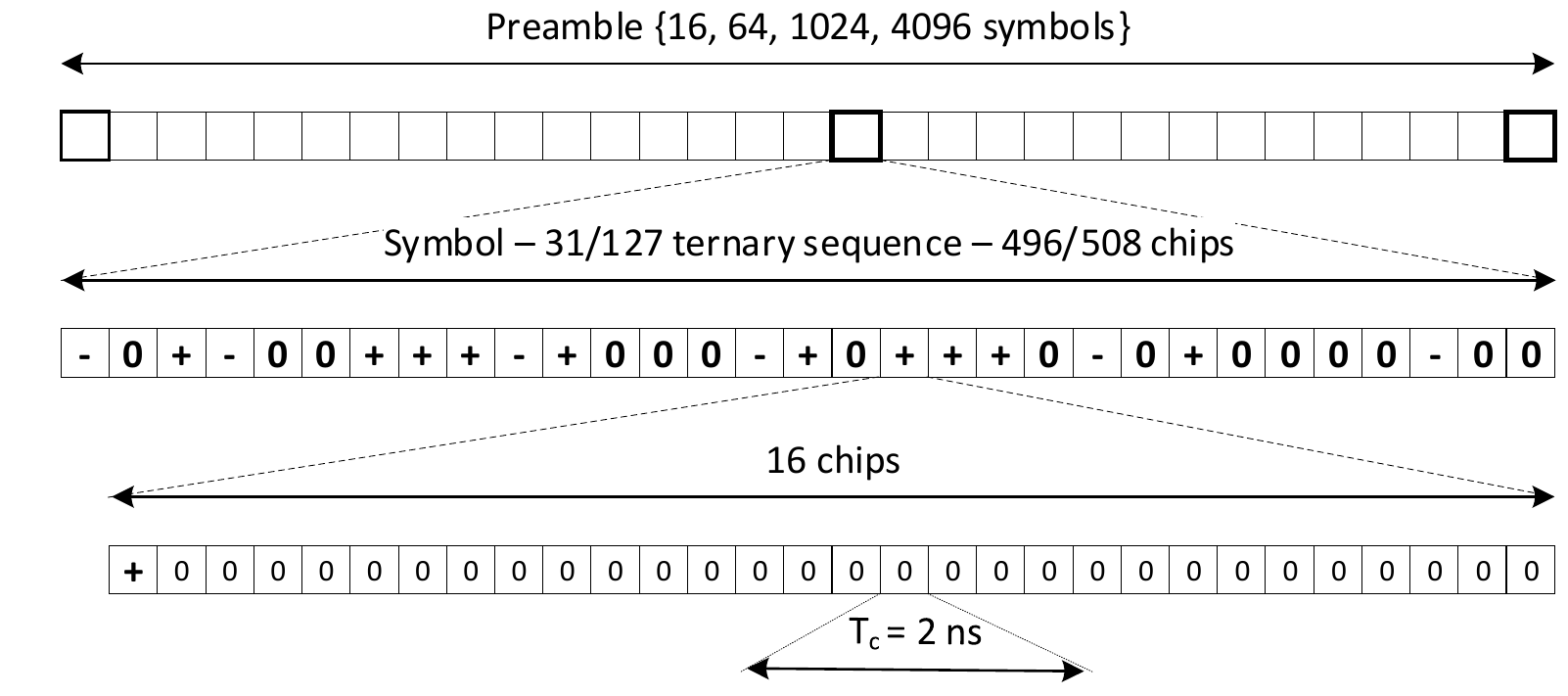}
\caption{IEEE 802.15.4a preamble.}
\label{fig:UWB_preamble}
\end{figure}

Burst position modulation (BPM) and binary phase shift keying (BPSK) are used to encode the transmitted information. The chip or pulse duration is about 2 ns, which corresponds to a bandwidth of 500~\!MHz. The CIR is estimated using the synchronization preamble. The synchronization preamble consists of 16, 64, 1024, or 4096 symbols. Each symbol of the preamble contains a ternary preamble code {-1,0,1} with a length of 31 or optionally 127. The code with a length of 31 is distributed with 16 chips, which a duration is approximately 2~\!ns. The code is transmitted only in the first chip. Similarly, the code with a length of 127 is spread over 4 chips. The IEEE 802.15.4a preamble is plotted in Figure~\ref{fig:UWB_preamble}. This preamble allows the CIR to be estimated for excess delays below the duration of a symbol, i.e., about 1 µs. For example, the Decawave chipset DW1000 calculates the 1016 complex samples CIR with a sampling rate of about 1~\!ns, which corresponds to the delay caused by a path difference between radio rays of 30~\!cm.

\section{Impact of channel variation rate on the estimated CSI}
\label{sec:Channel_variation}

In communication systems, the time over which the wireless channel is expected to be static is called the coherence time $T_c$. The coherence time is related to the Doppler spread, which is the result of the narrow band signal spectrum widening due to multipath propagation and the relative motion of the transmitter and receiver. The Doppler spread is represented as a Doppler spectrum bounded by the maximum Doppler frequency shift $f_m = \frac{v}{c} f_c$, where $c$ is the speed of light, $v$ is the terminal velocity, and $f_c$ is the carrier frequency. A simple approximation for the coherence time is
\[
T_c \approx \frac{1}{f_m}.
\]
This approximation indicates a time interval in which a Rayleigh fading signal can vary significantly. In reality, however, the CIR is random, so the coherence time should be considered a statistical measure.

In other words, the coherence time can be considered as a period of time during which the CIR is essentially invariant, i.e., the symbols transmitted during this period experience the same channel impairments. The coherence time can be defined as an interval in which the channel impulse autocorrelation function is above 0.5~\cite{Rappaport_1996}.
\[
T_c \approx \frac{9}{16 \pi f_m}.
\]
This definition is very restrictive so in practice, the geometric mean of both definitions is usually in use
\[
T_c \approx \sqrt {\frac{9}{16 \pi}} \frac{1}{f_m}.
\]

Coherence time for some selected LTE and NR bands and some typical user terminal speeds are shown in Table~\ref{tab:Chorence_time}. To ensure connectivity, the channel coherence time must be longer than the symbol or subframe length. The symbol time in LTE systems is 0.06667~\!ms, while the subframe length is 1.0~\!ms. In the LTE bands, the requirements are met up to the speed of the vehicle on the highways, while the LTE communication for fast trains may be faulty in highly scattered environments. The communications operating in the mm frequency band may exhibit problems even at lower user terminal speeds. From Table~\ref{tab:Chorence_time}, it can be seen that the channel characteristic is not constant for the entire duration of the subframe, which lasts 1~\!ms when the user terminal is in motion. In other words: only indoor communication may be based on the assumption that the radio channel does not significantly vary with time. The problem is less pronounced at NR, where symbols are even shorter at high-frequency bands, namely, 0.03333~\!ms, 01667~\!ms, 0.00833~\!ms, and 0.00417~\!ms, and the basic transmission element is the resource block, whose duration is only a single symbol. For Wi-Fi, the symbol duration is 0.004~\!ms, but the packets can be longer and reach 10~\!ms, leading to the conclusion that Wi-Fi can be used to estimate the channel to walking speed. Similar conclusions can be drawn for the various wireless sensor network technologies.

\begin{table} [!h]
\centering
\caption{Coherence time in ms for some LTE and NR bands and speed of user equipment.}
\label{tab:Chorence_time}
\resizebox{1.00\columnwidth}{!}{%
\begin{tabular}{r||c|c|c|c}
$f_c$ [MHz] & $v=4$ km/h & $v=60$ km/h & $v=120$ km/h  & $v=240$ km/h \\ [4pt] \hline \hline \tstrut
    0.7~~  & 163.20~~  &	10.88~~  &	5.44~~   &	 2.72   \\ [4pt]  \tstrut
    0.9~~  & 126.93~~  &	8.46~~   &	4.23~~   &	 2.12   \\ [4pt]  \tstrut
    1.8~~  & 63.47~~   & 4.23~~   &  2.12~~   &	1.06   \\ [4pt]  \tstrut
    2.1~~  & 54.40~~   &	3.63~~   &  1.81~~   &	0.91   \\ [4pt]  \tstrut
    2.6~~  & 43.94~~   &	2.93~~   &  1.46~~   &	0.73   \\ [4pt]  \tstrut
    3.6~~  & 31.73~~   &	2.12~~   &  1.06~~   &	0.53   \\ [4pt]  \tstrut
    5.9~~  & 19.36~~   &	1.29~~   &  0.65~~   &	0.32   \\ [4pt]  \tstrut
    28.0~~ &  4.08~~   &	0.27~~   &  0.14~~   &   0.07   \\ [4pt]  \tstrut
    47.0~~ & 2.43~~   &	0.16~~   &  0.08~~   &   0.04   \\ [4pt]
    \hline 
\end{tabular}%
}
\end{table}

\section{Discussion}
\label{sec:Discussion}
To benefit from the estimated CSI, the information must be stored in an intelligent database. The CSI can be represented as CIR  $h(t,\tau)$ or channel transfer functions $H(t,f)$ and CIR-derived parameters such as the attenuation or gain of the  channel, coherence bandwidth, coherence time, interference from other systems, i.e.  Furthermore, the information about the carrier frequency, the channel bandwidth, and measurement system must be stored for estimating the CSI. We are concerned with propagation characteristics that change slowly and therefore can be stored in a database for further use. This category includes indoor wireless channels, where the terminal is static or moving at low speed, and outdoor channels, but only considering shadowing effects and not the effects of local scattering, which can change when the terminal slightly changes its position or orientation.

Currently, only chipsets that obey IEEE 802.15.4 standard provide raw information about the CIR, but we expect this information to be available for other emerging wireless technologies. The channel representation of the DW1000 chipset contains 1016 samples of real and imaginary channel coefficients represented by 16 bits. The sampling time is 1~\!ns, which corresponds to a channel bandwidth of 500~\!MHz. The frequency resolution is thus
\[
\Delta f = \frac{1}{T N} = \frac{1}{1016 * 10^{-9} s} \approx 1 \text{MHz}.
\]

Similarly, we can calculate the CIR  for the LTE and NR, where the $\Delta f = 15$~\!kHz. The bandwidth, the number of resource blocks, the subcarrier bandwidth, the number of samples to represent the CIR, the sampling time, and the maximum excess delay are given in Table~\ref{tab:xG_BW_Delay_Spread}. The parameters given for LTE differ significantly from those of the UWB system due to the large differences in system bandwidths. The sampling intervals, column 6, are longer and, as expected, they decrease with bandwidth 

The LTE parameters are similar to the WiFi parameters. WiFi exploited 52~\!OFDM symbols to estimate the CIR. The OFDM symbol occupies a bandwidth of 313.2~\!kHz, giving an effective bandwidth of about 16 MHz and a sampling interval of 61.5~\!ns. The maximum excess delay covered corresponds to the maximum excess delay expected in LTE and LTE Advanced systems.

\begin{table} [!h]
\centering
\caption{LTE system: bandwidth, number of resource blocks, subcarrier bandwidth, number of samples for channel impulse representation, sampling time, and maximum excess delay.}
\label{tab:xG_BW_Delay_Spread}
\resizebox{1.00\columnwidth}{!}{%
\begin{tabular}{c|c|c|c|c|c|c}
	      & Number of & Subcarrier  & Effective    & Number of    &	Sampling & Maximum     \\
Bandwidth  	& resource  & bandwidth	  &	bandwidth    & samples for  & interval     & excess delay  \\
      MHz	  &	blocks    &	MHz	        & MHz	       & $h(\tau)$	  & ns       & ${\mu s}$	\\ [4pt]
             \hline \hline \tstrut
1.4		&	6	&	15	&	1.08	&	72		&	925.93	&	66.67	\\ [4pt]  \tstrut
3.0		&	15	&	15	&	2.70	&	180		&	370.37	&	66.67	\\ [4pt]  \tstrut
5.0		&	25	&	15	&	4.50	&	300		&	222.22	&	66.67	\\ [4pt]  \tstrut
10.0	&	50	&	15	&	9.00	&	600		&	111.11	&	66.67	\\ [4pt]  \tstrut
15.0	&	75	&	15	&	13.50	&	900		&	74.07	&	66.67	\\ [4pt]  \tstrut
20.0	&	100	&	15	&	18.00	&	1200	&	55.56	&	66.67	\\ [4pt]  \tstrut
50.0	&	250	&	15	&	45.00	&	3000	&	22.22	&	66.67	\\ [4pt] 
    \hline
\end{tabular}%
}
\end{table}

Contemporary communication systems are equipped with multiple antennas. To consider this aspect as well, the CIR should be estimated for each pair of antennas between the transmitter and receiver. NR and LTE support the estimation of CSI per pair of antennas by introducing the concept of antenna ports and defining the training signal in resource elements such that they do not interfere with each other.

Due to noise and interference, the communication systems provide the average value of the estimated CSI. To obtain complete information about CSI, the CSI statistical parameters should be estimated such us  probability density function or at least the standard deviation of the probability density function. Unfortunately, at the current stage of CSI estimation, no wireless communication system provides this data.

The parameters to be stored in the database must be identified. The wireless channel varies due to random variations in the environment around the transmitter and the receiver, the time variation of the interference, and the random movement and rotation of the transmitter and receiver. The information stored in the database consists of parameters that do not change significantly in time, such as the mean and standard deviation of the received signal strength, the CIR, etc. We have identified the following parameters to be stored in the database:
\begin{itemize}
    \item The power received from a particular base station, which is in the 4G and 5G communication system called Reference Signal Received Power (RSRP). In systems with limited channel estimation functionality, the parameter is rarely available.
    \item Received signal strength, which is composed of the received interference power, noise  in addition to the received power of the respective base station. In many systems, the parameter is referred to as the Received Signal Strength Indicator (RSSI). Due to its simple estimation, the RSSI is accessible as a parameter in almost all wireless communication systems.  
    \item Carrier frequency, the parameter is known or can be estimated from wireless communication system parameters.
    \item Channel bandwidth, the parameter is known from the wireless standard or, if configurable, it can be estimated from the beacons, broadcast from the base station.
    \item Channel impulse response (CIR) which includes:
    \begin{itemize}
        \item imaginary and real part of the CIR coefficients,
        \item sampling interval, i.e. time between samples. 
    \end{itemize}
    In early wireless communication systems, CIR is not estimated. In fact, for narrowband wireless communication systems, the attenuation of the wireless channel is sufficient for the operation of the system. In today's wireless communication systems, the CIR is estimated for the bands with an active connection. However, NR supports the estimation of CSI even for bands not occupied by an active connection.
\end{itemize}

In addition to the above parameters, some parameters related to the rate of channel variation would be beneficial, e.g., channel coherence time, Doppler frequency spread, and channel coherence bandwidth.

\section{Conclusions}
\label{sec:Conclusion}

This paper surveys the ability of terrestrial wireless communication systems to estimate channel state in terms of CIR, power delay profile, or wireless channel attenuation. While early wireless communication systems provide means for estimating narrowband wireless channel parameters such as channel attenuation, the next generation, including 2G, 3G, 4G, Wi-Fi, and some wireless sensor network technologies, is capable of estimating broadband wireless channel parameters. The estimated CSI, except for some basic ones such as RSRP, RSSI, are not available through defined and publicly available APIs, and the CSI is measured only for wireless channels with active communication. The NR provides the possibility to measure CSI also in wireless channels without active communication.

Only UWB wireless technology, based on the IEEE 802.15.4 (2011) standard, provides a good experimental platform for observing CSI and CIR and performing some combined sensing and communication experiments.

%

\bibliographystyle{IEEEtran}
\bibliography{references}

\end{document}